\documentclass[fleqn,10pt]{wlscirep}
\usepackage[utf8x]{inputenc}
\usepackage[T1]{fontenc}
\usepackage{braket}
\usepackage{amsmath}
\usepackage{amssymb}
\usepackage{verbatim}
\usepackage{cite}
\usepackage{hyperref}
\usepackage{nameref}
\usepackage{bm}
\newcommand{\bb}[1]{\mathbf{#1}}
\DeclareMathOperator{\diag}{diag}
\DeclareMathOperator{\iu}{i}
\usepackage{color}
\definecolor{Blue}{rgb}{0.3,0.3,0.9}
\definecolor{Red}{rgb}{0.9,0.3,0.3}
\definecolor{Green}{rgb}{0.3,0.6,0.3}
\newcommand{\revision}[1]{\textcolor{Red}{{#1}}}

\title{Spin-polarized localization in a magnetized chain}

\author[1,*]{Leonardo Benini}
\author[2]{Amrita Mukherjee}
\author[3,2]{Arunava Chakrabarti}
\author[1]{Rudolf A. R\"omer}
\affil[1]{Department of Physics,
University of Warwick, Coventry CV4 7AL, U.K.}
\affil[2]{Department of Physics, University of Kalyani, Kalyani, West Bengal-741 235, India}
\affil[3]{Department of Physics, Presidency University, 86/1, College Street, Kolkata 700073, West Bengal, India}

\affil[*]{l.benini@warwick.ac.uk}

\begin{abstract}
We investigate a simple tight-binding Hamiltonian to understand the stability of spin-polarized transport of states with an arbitrary spin content in the presence of disorder. The general spin state is made to pass through a linear chain of magnetic atoms, and the localization lengths are computed.  Depending on the value of spin, the chain of magnetic atoms unravels a hidden transverse dimensionality that can be exploited to engineer energy regimes where only a selected spin state is allowed to retain large localization lengths. We carry out a numerical anmalysis to understand the roles played by the spin projections in different energy regimes of the spectrum. For this purpose, we introduce a new measure, dubbed spin-resolved localization length. We study uncorrelated disorder in the potential profile offered by the magnetic substrate or in the orientations of the magnetic moments concerning a given direction in space. Our results show that the spin filtering effect is robust against weak disorder and hence the proposed system should be a good candidate model for experimental realizations of spin-selective transport.
\end{abstract}
\begin{document}

\flushbottom
\maketitle
%

\section*{Introduction}
The classical paradigm of charge transport in semiconductors, based on the propagation of electrons or holes, has been dressed, in recent years, with the novel concept of spin transport. The field of \emph{spintronics} \cite{Prinz1998,Zutic2004Spintronics:Applications} evolved rapidly in a couple of decades, bringing forward the idea of transporting information through the spin degrees of freedom of electrons, along with or even replacing the role of its charge. To allow such progress, a strong bond emerged between traditional solid state physics and materials science \cite{Bercioux2015QuantumReview}.
Spintronics exploits the ability of conduction electrons to carry a spin-polarized current \cite{Pareek2002SpinInteraction}, and relies heavily on long decoherence time and length scales, and leads to the elegance of multifunctionality, an increased data processing speed and less power consumption \cite{Matityahu2013SpinInterferometer}. It is pivotal in quantum information technology and has opened up the possibility of developing quantum computers. This vast potential is definitely what puts the subject of spin-polarized transport on the forefront of contemporary solid state and materials science research.

Experiments in this direction have been inspiring, especially in the last two decades. The development of molecular wires and spin-polarized tunneling devices \cite{Andres1996} helped to explore spin-polarized transport in low dimensional systems. Studies on the quantum confinement effect on electron transport \cite{Goldhaber-Gordon1998,Cronenwett1998,Yu2005SpinMode}, spin filtering and switching effects using ferromagnetic `leads' and external fields \cite{Frustaglia2001QuantumSwitch,Koga2002Spin-FilterDiode}, precision maneuvering over spin currents \cite{Lu2007PureInteraction} and other experiments on spin-polarized conductance in carbon nanotubes \cite{Sahoo2005ElectricalContacts,Alam2015SpinDNA} have enriched the understanding and, at the same time, have brought the scope of this relatively recent field of spin transport into limelight. Besides, studies on the spin selective properties of chiral double-stranded DNA molecules \cite{Gohler2011SpinDNA.,Xie2011SpinOligomers}, on the spin filtering effect exhibited by chiral molecules \cite{Mathew2014Non-magneticTemperature}, and spin injection and spin transport in organic spin valves \cite{Geng2016EngineeringBrushes} have widened the horizon of the subject. 
Comprehensive theoretical knowledge also accumulated over the years, not only corroborating the experimental findings but also posing challenges concerning the design of new experimental setups. A simple quantum interferometer, designed in the shape of a single mode ring and threaded by a magnetic flux \cite{Buttiker1984QuantumRings,Cheung1988PersistentRings}, has proven to be a well-suited object to understand the interplay of a closed loop geometry and a trapped magnetic flux in terms of the path-breaking Aharonov-Bohm (AB) effect \cite{Aharonov1959SignificanceTheory}. Such a simple system with a two-terminal lead geometry was later exploited  \cite{Wang2008AnisotropicInteractions} to study the anisotropic spin transport in the presence of the Rashba and Dresselhaus spin-orbit interactions \cite{Bychkov1984OscillatoryLayers,Dresselhaus1955Spin-OrbitStructures}. Earlier, a model quantum ring with a side-coupled stub was proposed for potential spin filters with comprehensive control \cite{Lee2006SpinWire}. Recently, an analytical tight binding model has been proposed to study the spin-polarized transport in DNA molecules as well \cite{Varela2016EffectiveTransport}.

In this context, it is worth pointing out that tight binding models of magnetic chains, acting as spin filters under various geometrical configurations, have provided a set of quite interesting tools in understanding certain subtle aspects of spin filtering. In this communication, we shall be focusing on such a model. A variety of such studies rely on describing a one-dimensional nano-wire by `magnetic' atoms whose magnetic moments interact with the spin of the projectile that is incident on the system from one side \cite{Shokri2005,Dey2011,Shokri2006,Mardaani2006}.
Very recently we brought out the conditions of spin filtering a periodic array of magnetic atoms, extending the previous concepts to deal with projectiles with higher spins \cite{Pal2016e}. The idea was later implemented to study the interplay of quasiperiodic arrangements and the topology of specific model networks comprising magnetic and non-magnetic atoms, having a minimal quasi-one dimensionality. It was shown how a hidden dimension in the linear chain of networks opened up, depending on the spin state of the incoming projectile \cite{Mukherjee2018Flux-drivenNetworks}, which could be exploited to filter out any desired spin channel. The correlations between the parameters of the Hamiltonian and an external magnetic field were found crucial in delocalizing almost all the single particle states over continuous zones in the energy spectrum, thus obtaining a clean spin filtering.

Spin-polarized transport studies of higher-spin atoms have not yet received the same attention as more conventional spin-$1/2$ electronic transport. However, large-spin atomic gases have already been the subject of promising theoretical investigations since the end of the last century \cite{Ho1999,Yip1999ZeroSpin,Wu2003,Honerkamp2004UltracoldModel}, mainly inspired by the discovery of Bose-Einstein condensation in atomic gases and the search for superfluid phases in alkali atoms. Fermionic alkaline-earth atoms with large nuclear spin have also been considered for quantum information processing and as quantum simulators for many-body exotic phenomena \cite{Gorshkov2010Two-orbitalAtoms}. In the last decade, great experimental progress in the field of ultracold atoms and optical lattices have made the realization of such systems possible. A far-from exhaustive list of examples include the study of quantum degeneracy in an ultracold gas of $^{87}$Sr atoms \cite{Desalvo2010}, the realization of a 1D gas of fermions with tunable spin \cite{Pagano2014ASpin} ,investigations on exotic forms of magnetism related to the SU$(N)$ symmetry emerging in cold atoms with high-spin states \cite{Zhang2014SpectroscopicMagnetism}, and the gas mixture of two fermionic isotopes of  Ytterbium, $^171$Yb (nuclear spin $I=1/2$) and $^173$Yb (nuclear spin $I=5/2$) \cite{Taie2010}. Theoretical \cite{Ho1998,Ciobanu2000,Zhou2001SpinCondensates, Demler2002SpinorFractionalization,Kuhn2012} and experimental \cite{Stenger1998SpinCondensates,Stamper-Kurn1999QuantumCondensate,Leanhardt2003} efforts in the study of bosonic gases have been equally successful\cite{Kawaguchi2012SpinorCondensates}, attracting an increasing amount of attention due to their rich variety of phenomena.

\begin{figure}[ht]
\centering
\includegraphics[width=0.5\columnwidth]{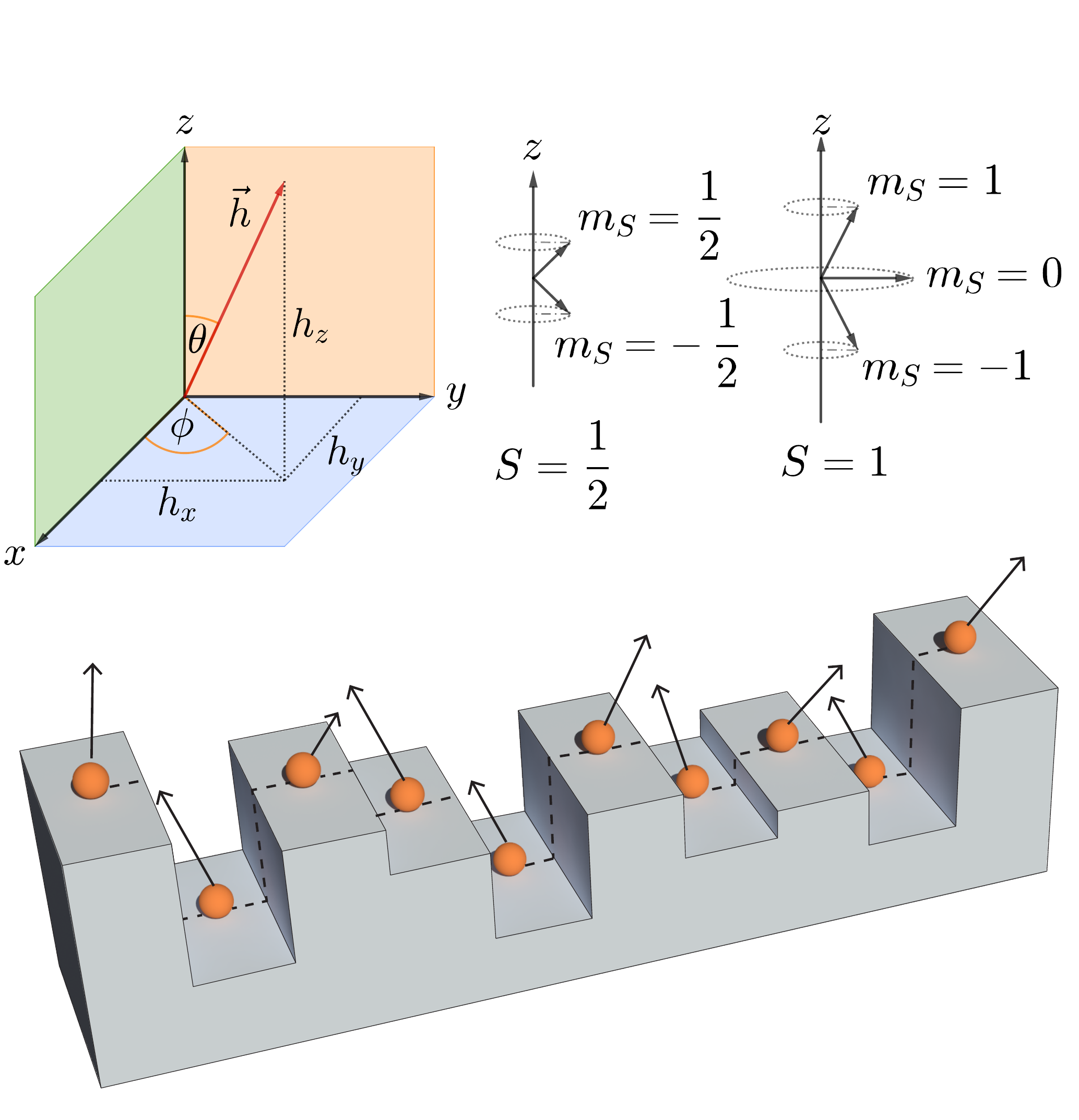}
\caption{(Color online). Top left: decomposition of magnetic moment $\vec{h}$ in its three-dimensional components $h_x, h_y, h_z$. $\theta$ and $\phi$ are called polar and azimuthal angles, respectively. Top right: graphic depiction of spin projections for two different spin states $S=1/2$ and $S=1$. Bottom: a schematic picture of a linear spin chain in the presence of potential and magnetic field disorder. The random heights at which the particles are positioned simulate the irregularity of the substrate surface and thus the disorder on the onsite potential. The random orientations of the field vector represent the disorder in the magnetic field.} 
\label{fig-schematic}
\end{figure}

In this paper, we study the effect of random, uncorrelated disorder on spin filtering for systems of spin-1/2 and spin-1 particles. We do this through an exhaustive calculation of the \emph{spin-resolved localization lengths}, specifically designed for our purpose, in various cases of disorder and employing a transfer matrix method (TMM). We use a tight binding model, similar to what we employed before \cite{Pal2016e,Mukherjee2018Flux-drivenNetworks}. A chain of magnetic atoms is considered and randomness is introduced in the values of the on-site potentials and in the magnitudes as well as in the orientations of the magnetic moments with respect to a preferred axis as shown in Fig.\ \ref{fig-schematic}.
Extensive numerical calculations of localization lengths reveal that, even in the presence of low to moderate disorder, different spin components can be selectively localized in different regimes of energy. While one spin component gets localized in a certain subband, the remaining ones can remain extended, at least within the system size considered. This implies that, spin filtering can indeed be observed at different subsections of the `energy bands', and the phenomenon of thus can be assessed, using the TMM, to be robust even in the presence of randomness. 
The robustness of the phenomenon may be inspiring for experimentalists in designing quantum wires where a bit of disorder can creep in, and yet the system can act as a stable spin filter.

\section*{Results}
\subsection*{\label{sec:model}Model}
We exploit a variation of a spin-polarized tight-binding Hamiltonian describing a linear chain of atoms fixed to a magnetic substrate supplying a magnetic moment $\vec{h}_n$ to the $n$-th atom, given by \cite{Shokri2005,Dey2011,Shokri2006,Mardaani2006}
\begin{equation}
\bb{H}=\sum_n \bb{c}^{\dag}_n \left(\bb{\epsilon}_n -\vec{h}_n \cdot {\bb{S}}_n \right) \bb{c}_n + \sum_{\langle n,m\rangle} \left(\bb{c}^{\dag}_n \bb{t}_{n,m} \bb{c}_m + 
h.c.\right),
\label{ham}
\end{equation}
where each of the entries- $\bb{c}^\dagger_n$, $\bb{c}_n$, $\bb{\epsilon}_n$, $\bb{t}_{n,m}$ and $\bb{S}_n$ are multi-component quantities expressed in terms of the spin projections $m_S=-S, -S+1, \cdots, S-1, S$, for a total of $2S+1$ components.

In the spin-1/2 case, the only available components can be labelled as $\uparrow, \downarrow$ and subsequently we have 
\begin{equation}
\mathbf{c}^{\dag}_n = \left(
c^{\dag}_{n,\uparrow} c^{\dag}_{n,\downarrow}  
\right), \qquad
\mathbf{c}_n = 
\left( 
c_{n,\uparrow}  \\ 
c_{n,\downarrow}  
\right), \qquad
\bb{\epsilon}_n = \diag(\epsilon_{n,\uparrow}, \epsilon_{n,\downarrow}), \qquad
\bb{t}_{n,m} = \diag(t, t),
\label{matrices}
\end{equation}
where $\bb{c}^\dagger_n$($\bb{c}_n$) is the creation (annihilation) operator at site $n$, $\bb{\epsilon}_n$ is the diagonal on-site energy and $t$ is the spin-independent, uniform hopping integral along the chain.
The energies $\epsilon_n$ are assumed to be independently picked random variables within a uniform distribution in $[-W/2, W/2]$, with $W$ determining the degree, or strength, of disorder in the system. 
If a particle is injected in the system, its spin interacts with the local magnetic moment $\vec{h}_n$ at site $n$ through the term $\vec{h}_n \cdot \bb{S}_n=h_{n,x}\bm{\sigma}_x +h_{n,y}\bm{\sigma}_y+ h_{n,z}\bm{\sigma}_z$, where $\bm{\sigma}_x, \bm{\sigma}_y, \bm{\sigma}_z$ are the Pauli spin matrices (or the generalized Pauli matrices $\bm{\sigma}^{(S)}$ for higher spins). Explicitly, in the spin-1/2 case we have
\begin{equation}
\vec{h}_n \cdot \bb{S}_n=
\left( \begin{array}{cccc}
h_n \cos{\theta_n} &  h_n \sin{\theta_n} e^{-i\phi_n}\\ 
h_n \sin{\theta_n} e^{i\phi_n} & -h_n \cos{\theta_n} 
\end{array}
\right) ,
\label{interact}
\end{equation}
with $\theta_n$ and $\phi_n$ the polar and azimuthal angles, respectively. We can write an analytical form of the Schr\"{o}dinger equation for the two different spin channels ($\uparrow$ and $\downarrow$) at energy $E$ starting from the general time-independent equation $\bm{H}\boldsymbol{\psi}=E\boldsymbol{\psi}$, with $\boldsymbol{\psi}=\{\psi_{1,\uparrow}, \psi_{1,\downarrow}, \ldots, \psi_{n,\sigma}, \ldots \}$. We note that the band width is $E_B= 4 t + 2h$ for all $S$ in the clean case $W=0$.
Using the Hamiltonian \eqref{ham} we get recursive equations relating the wave function at site $n$ with the neighboring at sites $n\pm 1$. When recast in the transfer matrix formalism, we find 
\begin{equation}
    \begin{pmatrix}
    \psi_{n+1,\uparrow} \\ \psi_{n+1,\downarrow} \\ \psi_{n,\uparrow} \\ \psi_{n, \downarrow}
    \end{pmatrix}
    =\begin{pmatrix}
    \frac{(E-\epsilon_{n,\uparrow}+h_n\cos\theta_n)}{t} & \frac{h_n\sin\theta_n}{t} & -1 & 0 \\ 
    \frac{h_n\sin\theta_n}{t} & \frac{(E-\epsilon_{n,\downarrow}-h_n\cos\theta_n)}{t} & 0 & -1 \\
    1 & 0 & 0 & 0 \\
    0 & 1 & 0 & 0
    \end{pmatrix}
   \begin{pmatrix}
    \psi_{n,\uparrow} \\ \psi_{n,\downarrow} \\ \psi_{n-1,\uparrow} \\ \psi_{n-1, \downarrow}
    \end{pmatrix}=\mathbf{T}_n \begin{pmatrix}
    \psi_{n,\uparrow} \\ \psi_{n,\downarrow} \\ \psi_{n-1,\uparrow} \\ \psi_{n-1, \downarrow}
    \end{pmatrix},
    \label{receq}
\end{equation}
where $\mathbf{T}_n$ is the transfer matrix at site $n$ and we set $\phi=0$ from now on without loss of generality. Thus, we obtain the transfer matrix for the entire linear system as $\mathbf{T_L}=\prod_{n=1}^L\mathbf{T}_n$.
For spin-1 particles, the generalized spin matrices are
\begin{equation}
\bm{\sigma}^{(S=1)}_x=\frac{1}{\sqrt{2}}
	\begin{pmatrix}
   	0 & 1 & 0 \\ 1 & 0 & 1 \\ 0 & 1 & 0
    \end{pmatrix}, \qquad
\bm{\sigma}^{(S=1)}_y=\frac{1}{\sqrt{2}i}
	\begin{pmatrix}
   	0 & 1 & 0 \\ -1 & 0 & 1 \\ 0 & -1 & 0
    \end{pmatrix}, \qquad
\bm{\sigma}^{(S=1)}_z=
	\begin{pmatrix}
   	1 & 0 & 0 \\ 0 & 0 & 0 \\ 0 & 0 & -1
    \end{pmatrix},
\end{equation}
and the interaction term is given by
\begin{equation}
	\vec{h}_n \cdot \bb{S}^{(S=1)}_n= \begin{pmatrix}
   	h_n\cos\theta_n & \frac{h_n\sin\theta_n}{\sqrt{2}} & 0 \\ \frac{h_n\sin\theta_n}{\sqrt{2}} & 0 & \frac{h_n\sin\theta_n}{\sqrt{2}} \\ 0 & \frac{h_n\sin\theta_n}{\sqrt{2}} & -h_n\cos\theta_n
    \end{pmatrix}.
\end{equation} Thus, the transfer matrix at the $n$-th site for spin-1 particles reads as 
\begin{equation}
\label{schrod}
    \mathbf{T}^{(S=1)}_n=
    \begin{pmatrix}
    \frac{(E-\epsilon_{n,1}+h_n\cos\theta_n)}{t} & \frac{h_n\sin\theta_n}{\sqrt{2}t} & 0 & -1 & 0 & 0 \\
    \frac{h_n\sin\theta_n}{\sqrt{2}t} & \frac{(E-\epsilon_{n,0})}{t} & \frac{h_n\sin\theta_n}{\sqrt{2}t} & 0 & -1 & 0 \\
    0 & \frac{h_n\sin\theta_n}{\sqrt{2}t} & \frac{(E-\epsilon_{n,-1}-h_n\cos\theta_n)}{t} & 0 & 0 & -1 \\
    1 & 0 & 0 & 0 & 0 & 0 \\
    0 & 1 & 0 & 0 & 0 & 0 \\
    0 & 0 & 1 & 0 & 0 & 0 \\
    \end{pmatrix}.
\end{equation}
It is worth noting the hidden analogy between the one-dimensional model \eqref{ham} and the relative equations \eqref{receq} with the equations for spinless fermions on a two-strand ladder network \cite{Sil2008a}: the system can be imagined as two identical chains coupled laterally, with an effective random onsite potential $\epsilon_{n,\uparrow}-h_n\cos{\theta_n} $ for the virtual ``upper" strand corresponding to the $\uparrow$ component, and $\epsilon_{n, \downarrow}+h_n\cos{\theta_n}$ for the ``lower" strand representing the $\downarrow$ projection. From this point of view, the hopping $t$ is the independent hopping along individual strands in the network, while the off-diagonal terms $h_n\sin{\theta_n}$ turn on inter-strand hopping channels. This kind of mapping can be generalized to arbitrary spin values, with the number of strands corresponding to the $2S+1$ spin projections. Intuitively, this ladder picture could represent a more immediate way of visualizing the spin filtering phenomenon, imagining that under certain tuning procedures transport is allowed only in one (or some, for $S>1/2$) strand. 

It has been pointed\cite{Pal2016e} out that the clean chain could act like a quantum device opening a transmission window only for specific spin components depending on the chosen energy range, filtering out all the others. Interestingly, the study was able to show the same behavior for arbitrary spin particles ($S=1/2$, $1$, $3/2$, etc.). Our aim here is to prove the robustness of this filtering effect adding randomness to the system, using a new variation of the TMM algorithm able to extract the localization length for the various transport channels, discerning the individual contribution of each spin component.

\subsection*{Density of States}

As already stated elsewhere \cite{Pal2016e}, the tuning of $h_n$ is sufficient for the engineering of the energy bands corresponding to the various spin components, in the absence of magnetic flux. For the sake of simplicity, we restrict ourselves to the $S=1/2$ case, but the following discussion also holds for particles with higher spin. 

In the simplest case of $h_n=h$, $\theta_n=\theta=0$, the cross terms determining the hybridization of the spin channels in Eq. \eqref{receq} are canceled, and all moments will be parallel to each other, lying along the $z$-axis. In the limit of $t\rightarrow 0$, the energy spectrum will show sharp peaks at $E=\epsilon\pm h$, and the DOS will correspondingly exhibit two $\delta$-function-like spikes at these energy values. Switching on the hopping $t$ results in a broadening of the spikes, ultimately merging in subbands when $t \sim h$. Hence, for a given value of $\theta$, $h$ can be tuned to open or close a gap in the energy spectrum. 

We computed the local density of states (LDOS) for the $\uparrow$ and $\downarrow$ components evaluating the Green's function $\mathbf{G}=(E \mathbf{I}-\mathbf{H})^{-1}$ in the basis $\ket{n, \uparrow(\downarrow)}$. The LDOS is given as\cite{Pal2016e}
\begin{equation}
    \rho_{\uparrow \uparrow (\downarrow \downarrow)}=\lim_{\eta\rightarrow 0}\braket{n, \uparrow (\downarrow)|\mathbf{G(E+\iu \eta)| n, \uparrow (\downarrow)} }
\end{equation}
and the energy gap has the very simple form
\begin{equation}
    \Delta=\left\{ 
    \begin{array}{cc}
    \frac{h}{S}-4t & \text{for } \frac{h}{S}> 4t, \\
    0              & \text{otherwise.} 
    \end{array}
    \right.
    \label{eq:gapeq}
\end{equation}
This gap equation implies the existence of the critical magnetic strength $h_c= 4t S$ at which the bands meet, independently of the choice of $\theta$.

\subsection*{Spin-resolved localization length}

For general $S$, $\mathbf{T}_n$ is a matrix of size $2 (2S+1) \times 2 (2S+1)$ with $2S+1$ associated Lyapunov exponents. The inverse of the smallest Lyapunov exponent determines a localization length $\lambda$ \cite{Kramer1993Localization:Experiment}. 
\revision{This $\lambda$ does not distinguish the spin projections contributing to its value (see Methods for more detail). However, this is required to establish the possibility of spin selectivity.
In order to characterize which of the $2S+1$ spin channels is responsible for $\lambda$, we therefore introduce instead the \emph{spin-resolved  localization lengths}, namely,}
\begin{equation}
\Lambda_{\sigma}(n) = \sum_{m_S=1}^{2S+1} \lambda_{m_S}(n)
 \left( \left|\psi^{(m_S)}_{n,\sigma}\right|^2 + \left|\psi^{(m_S)}_{n-1,\sigma}\right|^2 \right),
 \label{eq:spinresolvedloclength}
\end{equation}
for $\sigma= -S, -S+1, \ldots, S$. 
\revision{In Eq.\ \eqref{eq:spinresolvedloclength}, the index $m_S$ spans all possible spin projections.
The value of $\Lambda_{\sigma}$ sums the contributions to the spin $\sigma$ coming from the $m_S$ different channels. Just as for the usual localization length $\lambda_{m_S}$, small values of $\Lambda_{\sigma} \sim 1$ indicate strong localization while larger values $\Lambda_{\sigma} \gg  1$ denote less localization.}
In the following we shall use $\Lambda_{\sigma}$ to indicate \emph{converged} estimates for $n \rightarrow\infty$, using the usual TMM convergence criteria \cite{Kramer1993Localization:Experiment} for the $\lambda_{m_S}$'s. For all the calculations, we have set an accuracy threshold of $99.9\%$ (error of the mean at most $0.1\%$ for each $\lambda_{m_S}$), corresponding to a system of the order of $n\sim 10^{10}$ sites in the centre of the energy range. 
%
Due to normalization of the $\psi_{n,\sigma}^{(m_S)}$, we note that $\Lambda_{\sigma} \leq \max(\lambda_{m_S}, m_S=1, \ldots, 2S+1)$ for all $\sigma$.
Thus, although the $\Lambda_{\sigma}$ are not the direct results of the TMM calculations, they are a proper measure of the localization properties in different spin transport channels.
In the following, we shall drop the more precise labeling of spin-resolved localization lengths for the $\Lambda_{\sigma}$ and merely call them localization lengths.

\subsection*{\label{sec:spinhalf}The spin-1/2 case}

As mentioned in the introduction, we are interested in establishing a spin-filtering that remains robust in the presence of disorder. In order to do so, we shall choose a small but not negligible onsite disorder  $W=0.5 t \ll E_B$. Also, we shall investigate how spatial and random variations in $h$ and $\theta$ affect the filtering. In all these cases, we shall take care to retain the connection to the clean case. Finally, we shall also study the drop in $\Lambda_{\sigma}$ when further increasing disorder. The energy scale is set by $t=1$.

We obtain the localization lengths $\Lambda_\uparrow$, $\Lambda_\downarrow$ and the DOS $\rho_\uparrow$, $\rho_\downarrow$ for two different band structure cases: (i) for $h=1$, we have $\Delta=0$ and the spectrum shows for $-1\le E\le 1$ an overlapping of the spin-$\uparrow$ and spin-$\downarrow$ bands. (ii) The model develops a gap with growing $h$, and we chose to set $h=3$ as the second case of study, representing a clear gap $\Delta= 2$ in the range $-1\le E\le 1$. We also present, in every case, the DOS for the corresponding `clean' system without disorder. \revision{This is for comparison with the disordered case in order to indicate that the energy range of non-zero localization lengths remains largely in agreement with the clean band structure for the chosen values of $W$.}

\subsubsection*{\label{sec:spinhalf-energy}Energy dependence}

\begin{figure}[tb]
\begin{center}
\includegraphics[width=\columnwidth]{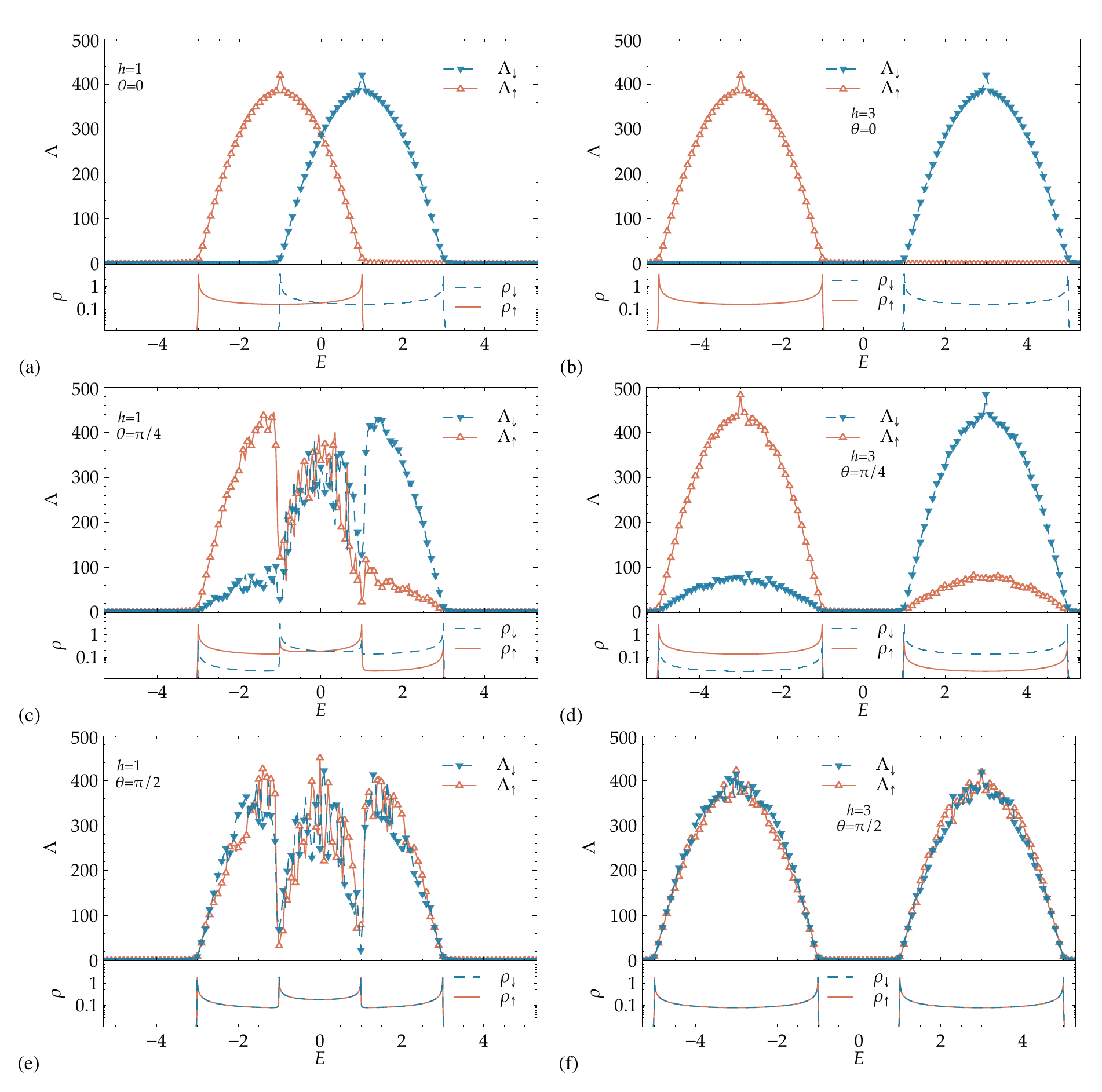}
\caption{(Color online) Projected localization lengths $\Lambda_{\sigma}=\lambda_{\sigma} |\Psi^2|$ for the disordered magnetic chain and DOS $\rho$ for the corresponding pure systems for spin-$1/2$ particles, with (a,c,e) showing overlapping bands at $h=1$ and (b,d,f), the well-separated bands at $h=3$. The value of the polar angle $\theta$ changes from (a,b) $\theta=0$ to (c,d) $\pi/4$ and (e,f) $\pi/2$. The strength of the potential disorder is $W=0.5$ in all cases. For clarity, only every second data point is indicated by a symbol, i.e.\ $\triangledown$ for spin-$\downarrow$ and $\triangle$ for spin-$\uparrow$. The DOS for the corresponding pure systems are shown in the lower part of the panels, corresponding to $\rho_\downarrow$ (dashed) and $\rho_\uparrow$ (solid line) for spin-$\downarrow$ and spin-$\uparrow$. Lines for $\Lambda_\sigma$ are guides to the eye. The $\lambda_\sigma$ error bars are within symbol size when not visible.} 
\label{fig-spinhalf-locpsi-dos}
\end{center}
\end{figure}

We show in the upper panels of Fig.\ \ref{fig-spinhalf-locpsi-dos} (a+b) the localization length for the case of weak disorder in the on-site potential and $\theta=0$. Since the off-diagonal terms are zero, we have complete channel separation, i.e.\ two uncoupled chains. The extent of the non-vanishing $\Lambda_\sigma$ values is in good agreement with the span of the DOS of the corresponding `clean' system as shown in the DOS plots in the lower panels of Fig.\ \ref{fig-spinhalf-locpsi-dos} (a+b).
We envision that the system obviously exhibits a clear separation into two distinct energy bands for spin-$\uparrow$ and spin-$\downarrow$. This spin filtering effect has already been discussed for the clean version of the system described above\cite{Pal2016e}. The effect survives the introduction of randomness in the distribution of the on-site potential: in Fig.\ \ref{fig-spinhalf-locpsi-dos} (a), for $-3\le E\le -1$,  only spin-$\uparrow$ electrons get transported through the chain, while they are blocked for $1 \le E \le 3$. Conversely, for $1\le E\le 3$, only spin-$\downarrow$ electrons contribute to transport and are blocked for $-3 \le E \le -1$.
As seen here, the subbands overlap in the range $-1 \le E \le 1$, preventing perfect filtering, but allowing a \emph{partial} filtering depending on the value of $E$.
In Fig.\ \ref{fig-spinhalf-locpsi-dos} (b), on the other hand, an obvious gap opens between the subbands, and the spin filtering is complete: only spin-$\uparrow$ electrons get transported in the lower subband and only spin-$\downarrow$ electrons transmit in the upper one. 

Of course, one has to be careful when speaking of ``transport''. Indeed, the finite localization lengths  indicate strong localization for any system much larger than the $\Lambda_\uparrow$, $\Lambda_\downarrow$ values found here. Nevertheless, the values of $\Lambda_\uparrow$, $\Lambda_\downarrow$ are of course determined by which value of $W$ was chosen to simulate a disordered environment. 
\revision{Let us assume for a moment that our system would have a finite length $L$ (say, e.g., $L=200$ in Fig.\ \ref{fig-spinhalf-locpsi-dos} (a+b)). The observation that we can find broad energy regions with either $\Lambda_\uparrow, \Lambda_\downarrow\gtrsim L$ clearly shows the robustness of the spin filtering to the presence of $W$. Indeed, were we to compute a \emph{spin-resolved} typical, dimensionless conductance by adapting a celebrated relation due to Pichard \cite{Pichard1984}, i.e.\
\begin{equation}
G_{\sigma}(n) = 
\sum_{m_S=1}^{2S+1} \frac{1}{\cosh^2{(L/\lambda_{m_S}(n))}}
 \left( \left|\psi^{(m_S)}_{n,\sigma}\right|^2 + \left|\psi^{(m_S)}_{n-1,\sigma}\right|^2 \right),
 \label{eq:spinresolvedconductance}
\end{equation}
we would find $G_\sigma\sim 1
$ for energy regions in which $\Lambda_\sigma\gtrsim L$. Conversely, $G_\sigma\sim 0$ when $\Lambda_\sigma < L$.}
Last, the spikes at $E=-1$ for $\Lambda_\uparrow$, and at $E=1$ for $\Lambda_\downarrow$ \revision{visible in Fig.\ \ref{fig-spinhalf-locpsi-dos} (a+b)} are due to the famous Kappus-Wegner degeneracy in 1D Anderson models at the middle of the bands \cite{Kappus1981AnomalyModel}.

Switching on the additional hopping between spin-$\uparrow$ and spin-$\downarrow$ channels for finite $\theta$ has a strong impact on spin filtering. Fig.\ \ref{fig-spinhalf-locpsi-dos} (c+d) show the results for $\theta=\pi /4$, for the same values of $h$ discussed above. One can easily see that \emph{mixing} between spin-$\uparrow$ and spin-$\downarrow$ states emerges. Looking at panel (d) for the sake of clarity, spin filtering still exists, but in a weaker sense: in the lower band, the spin-$\downarrow$'s can travel a distance lower than spin-$\uparrow$ particles by a factor of $\sim 5$, resulting in a more pronounced transport of the latter ones. 
\begin{figure}[tb]
\begin{center}
\includegraphics[width=\columnwidth]{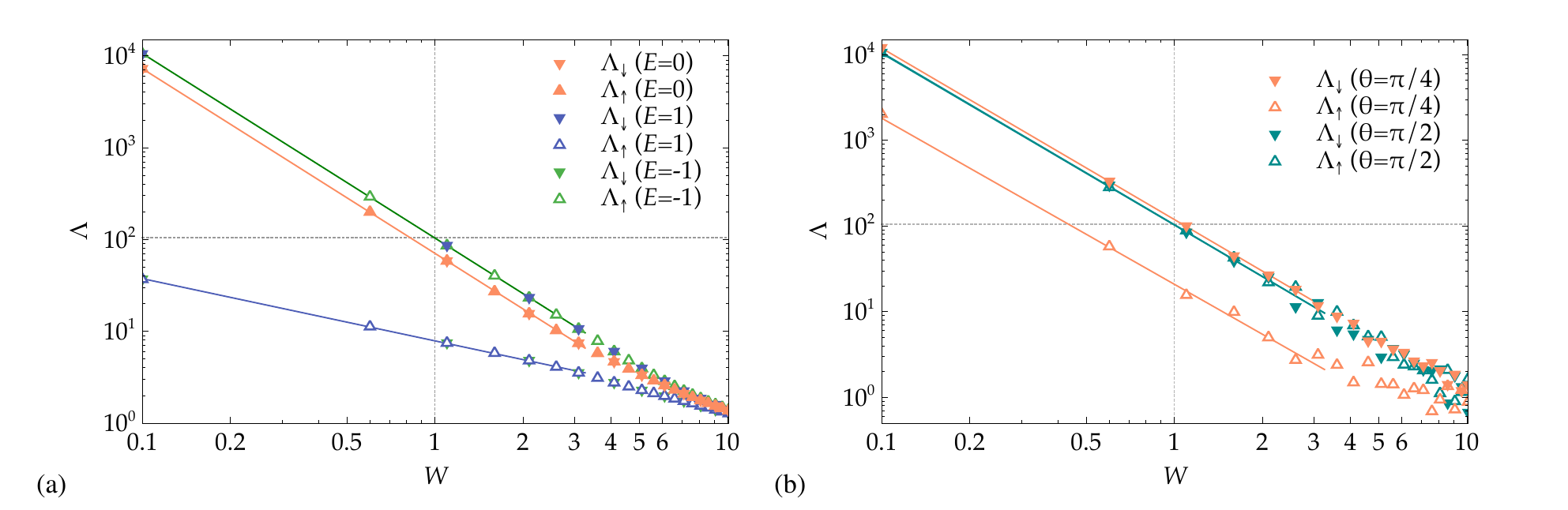}
\caption{(Color online) Disorder dependence of spin-resolved localization lengths $\Lambda_\sigma$ with (a) overlapping bands at $h=1$ for $\theta=0$ and (b) well-separated bands at $h=3$ for $\theta=\pi/2, \pi/4$. In (a) different colors correspond to different energy values, $E=-1$ (center of the lower subband), $E=0$ (bands crossing point) and $E=1$ (center of the higher subband). In (b) the plots show the results for the center of the $E=3$ subband, for different values of the polar angle $\theta$. In all panels, the solid lines show fits with Eq.\ \eqref{thouless} up to $W=4$, and the crossing point between the horizontal and the vertical dotted lines represents the band center Kappus-Wegner estimate $\lambda(W=1)=104 \pm 1$ \cite{Kappus1981AnomalyModel}.} 
\label{fig-spinhalf-locpsi-disorder}
\end{center}
\end{figure}
In panel (c) of Fig.\ \ref{fig-spinhalf-locpsi-dos}, we find that the spin-filtering for $-1\leq E \leq 1$ has stopped and neither spin-$\uparrow$ nor spin-$\downarrow$ particles dominate. The resulting ``transport'' will be spin-unpolarized.
Spin filtering disappears for all energies at $\theta=\pi/2$ as expected. Each spin component can be transported through the sample for the same distance for both subbands. The vanishing of spin filtering becomes evident by looking at the complete overlapping of both the $\Lambda_\uparrow$, $\Lambda_\downarrow$ localization lengths and $\rho_\uparrow$, $\rho_\downarrow$ DOS curves.
We can understand the contributions coming from the individual spin projections at finite $\theta$ by noting that an SU(2) rotation results in a decoupled basis $\boldsymbol{\Phi}_n=\mathcal{R}_\mathrm{SU(2)}^{-1} \boldsymbol{\psi}_n$  and the relations between the new basis vectors $\{\phi_{n,+},\phi_{n,-}\}$ and the old $\{\psi_{n,\uparrow}, \psi_{n,\downarrow}\}$ are given by
\begin{equation}
\label{eq-su2trafo}
\phi_{n,+}=\cos(\theta/2) \psi_{n,\uparrow}+ \sin(\theta/2) \psi_{n,\downarrow}, \qquad
\phi_{n,-}= - \sin(\theta/2) \psi_{n,\uparrow} + \cos(\theta/2) \psi_{n,\downarrow} .
\end{equation}
Here, $+$ and $-$ denote the projected spin components after the SU(2) rotation.
It is now easy to see that for, e.g., $\theta=\pi/2$ both states $\phi_{n,+}$ and $\phi_{n,-}$ have equally weighted contributions from the old $\psi_{n,\uparrow}$, $\psi_{n,\downarrow}$. 
The energy bands arising from the transformation \eqref{eq-su2trafo} are centered at $\epsilon-h$ and $\epsilon+h$, and they span the ranges $[\epsilon-h-2t, \epsilon-h+2t]$ and $[\epsilon+h-2t,\epsilon+h+2t]$, respectively. 
Obviously, the numerical results of Fig.\ \ref{fig-spinhalf-locpsi-dos} show exactly this behavior.

Before leaving this discussion of the spin-$1/2$ case, let us comment on the $\theta$ dependence of $\Lambda_{\sigma}$. Comparing Fig.\ \ref{fig-spinhalf-locpsi-disorder} (a+c+e), we see that there is a small increase of the maximal $\Lambda$ values when $\theta$ increases from $0$ and $\pi/4$ and$\pi/2$. In these latter situations --- including also Fig.\ \ref{fig-spinhalf-locpsi-disorder} (b+d+f) ---, we are effectively dealing with a coupled quasi-1D system, and it is well-known that this leads to an increase in the localization lengths \cite{ROMER2004}.


\subsubsection*{\label{sec:spinhalf-disorder}Disorder dependence}

We now study the disorder dependence $\Lambda_\sigma(W)$ at fixed values of $E$. We focus on some of the most significant energy values in the spectrum, and we present the results in Fig.\ \ref{fig-spinhalf-locpsi-disorder}.
For the gapless case $h=1$, we study the behavior of $\Lambda_\sigma$ for $\theta=0$ corresponding to the center of the lower subband at $E=-1$, the contact point between the subbands at $E=0$, and the center of the higher subband at $E=1$. We find that the standard perturbative small$-W$ Thouless fitting \cite{Thouless1972ASystems,Czycholl1981ConductivityResults,Kappus1981AnomalyModel}
\begin{equation}
\label{thouless}
\Lambda(W)=\mathcal{A}\ {W^{-\kappa}},
\end{equation}
works very well as shown in Table \ref{fittable}. We note that for majority states at the center of spin-split bands, even the Kappus-Wegner singularity with $\mathcal{A}= 104\pm 1$ is well recovered \cite{Kappus1981AnomalyModel}. For the mixed states at $E=0$, the value of $\mathcal{A}$ changes but $\kappa\approx 2$ remains valid. In the tails of the subbands, the  behavior is quantitatively different.   While the power-law \eqref{thouless} still holds, the $\mathcal{A}$ and $\kappa$ values have changed.
\begin{table}[tb]
\centering
\caption{Fit parameters for the $\Lambda_{\sigma}(W)$ plots in Fig.\ \ref{fig-spinhalf-locpsi-disorder} using Eq.\ \eqref{thouless}.}
\label{fittable1}
\label{fittable2}
\label{fittable}
\begin{tabular}{cccrcc}\hline\hline
$\Lambda$            & $h$ & $\theta$ & $E$ & $\mathcal{A}$ & $\kappa$ \\ \hline
$\Lambda_\downarrow$ & $1$ & $0$ & $0$  & $70.4 \pm 1.3$  & $2.01 \pm 0.02$ \\
$\Lambda_\uparrow$   & $1$ & $0$ & $0$  & $70.5 \pm 1.3$  & $2.01 \pm 0.02$ \\
$\Lambda_\downarrow$ & $1$ & $0$ & $1$  & $103.4 \pm 1.9$ & $2.01 \pm 0.02$ \\ 
$\Lambda_\uparrow$   & $1$ & $0$ & $1$ & $7.9 \pm 0.2$   & $0.68 \pm 0.02$ \\
$\Lambda_\downarrow$ & $1$ & $0$ & $-1$ & $7.9 \pm 0.2$   & $0.68 \pm 0.02$ \\
$\Lambda_\uparrow$   & $1$ & $0$ & $-1$ & $103.4 \pm 1.9$ & $2.01 \pm 0.02$ \\
$\Lambda_\downarrow$ & $3$ & $\pi/4$ & $3$ & $118.9\pm 0.1$  & $2.0003 \pm 0.0007$ \\
$\Lambda_\uparrow$   & $3$ & $\pi/4$ & $3$ & $20.87 \pm 0.01$  & $1.9384 \pm 0.0006$ \\
$\Lambda_\downarrow$ & $3$ & $\pi/2$ & $3$ & $103.7 \pm 0.2$   & $2.008 \pm 0.001$ \\
$\Lambda_\uparrow$   & $3$ & $\pi/2$ & $3$ & $102.64 \pm 0.01$ & $2.0004 \pm 0.0001$ \\ \hline\hline
\end{tabular}
\end{table}

Fig.\ \ref{fig-spinhalf-locpsi-disorder} (b) shows the behavior of $\Lambda_\sigma(W)$ for the gapped systems at $h=3$ for $\theta=\pi/4$, $\pi/2$ in the center of the higher subband at $E=3$. There is again good agreement with \eqref{thouless} for the $\kappa$ value. Since we are now effectively dealing with two coupled chains, the value of $\mathcal{A}$ can change\cite{ROMER2004}. Again, this is clear from the results of Table \ref{fittable}.

\subsection*{\label{sec:spinone}The spin-1 case}
\begin{figure}[tb]
\begin{center}
\includegraphics[width=\columnwidth]{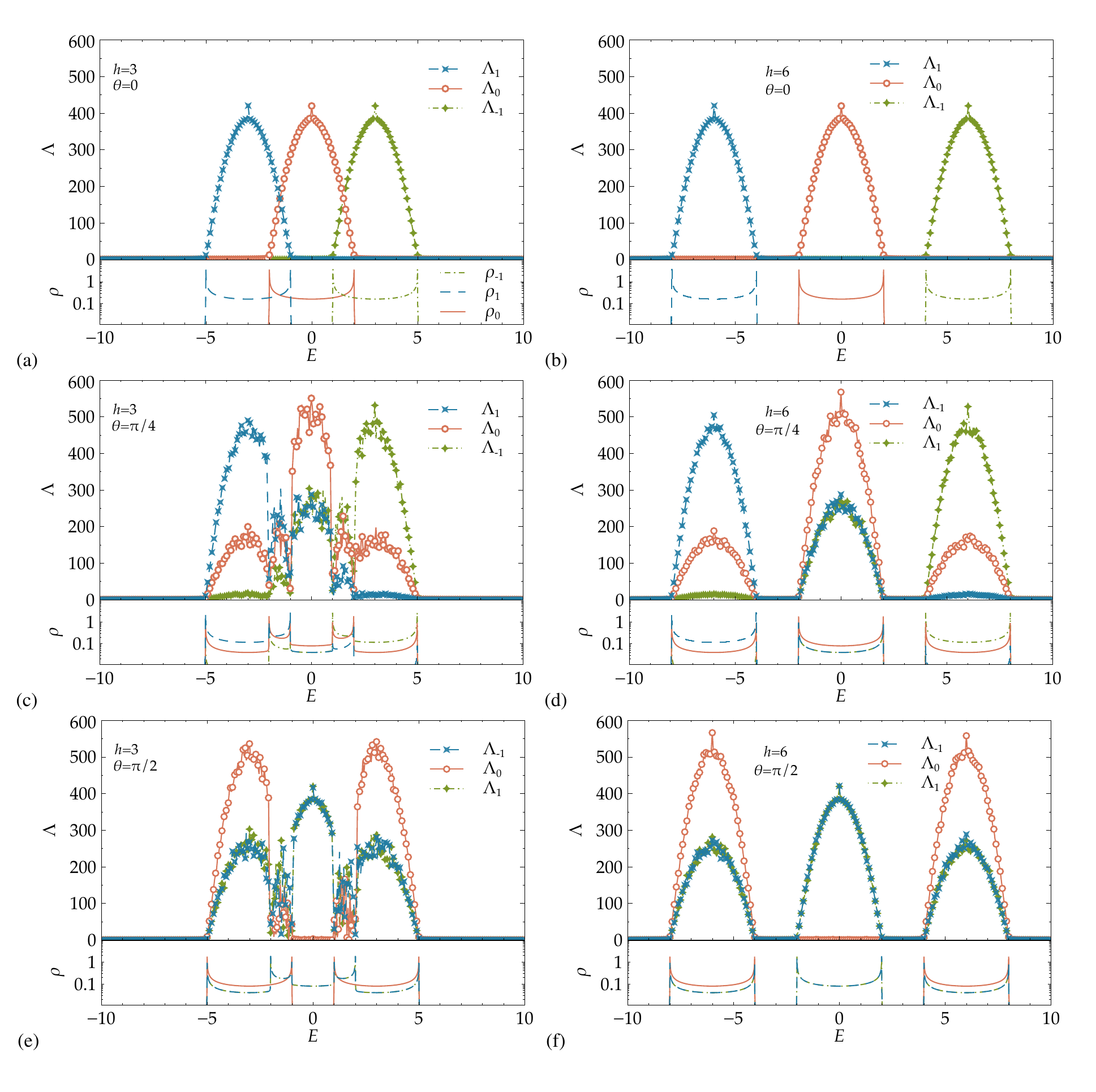}
\caption{(Color online) Projected localization lengths $\Lambda_{-1}$, $\Lambda_{0}$, $\Lambda_{1}$ and DOS for spin-$1$ particles with (a,c,e) overlapping bands at $h=3$ and (b,d,f) well-separated bands at $h=6$. The value of the polar angle $\theta$ changes from (a,b) $\theta=0$ to (c,d) $\pi/4$ and (e,f) $\pi/2$. As in Fig.\ \ref{fig-spinhalf-locpsi-dos}, we have $W=0.5$. Symbols and lines for $\Lambda_\sigma$ and $\rho_\sigma$ are as given in the figure legend. Only every second $\Lambda_\sigma$ symbol is shown for clarity. Lines for $\Lambda_\sigma$ are guides to the eye. The $\lambda_\sigma$ error bars are within symbol size when not visible.} 
\label{fig-spinone-locpsi-dos}
\end{center}
\end{figure}

For $\theta=0$, it can be seen in Fig.\ \ref{fig-spinone-locpsi-dos} (a+b) that the gap appears at higher values of the magnetic strength $h$, as we expect from Eq.\ \eqref{eq:gapeq}. 
For $h=3$ the subbands corresponding to spin components $m_S=-1$, $0$, $1$ overlap, and we have a clear gap for $h=6$. As in the previous $S=1/2$ scenario, there is perfect filtering for well-separated energy bands, while a partial filtering can be obtained for overlapping bands. 
The situation becomes quite complex for non zero values of $\theta$, as all three spin channels are coupled. We first again notice the slight increase in maximal $\Lambda_\sigma$ values due to the quasi-1D effect.
Next, the $m_S=0$ states have finite $\Lambda_\sigma$ in all three subbands for $\theta=\pi/4$ (panels (c) and (d)), while $m_S=-1$ spin states have none-zero $\Lambda_\sigma$ values only in the lower and central band and $m_S=1$ particles only in the upper and central ones. It is worth noting that all components have  none-zero $\Lambda_\sigma$ values in the central band $-2 \leq E \leq 2$, but the $m_S=0$ particle states are less localized. Thus, by selecting a system with finite size L comprised between 300 and 500, one can suppress the `-1' and `1' spin states while only the `0' spin-projection survives  and contributes to transport in the centre of the band.

A significant difference from the spin-$1/2$ states appears for spin-$1$ and $\theta=\pi/2$. Instead of complete mixing of the $\Lambda_\sigma$'s and the DOS curves, we observe that the $m_S=0$ states spread entirely into the left and right subbands, while they vanish in the central band. An argument analog to the SU(2) rotation for $S=1/2$ in the context of the band mixing given in the previous section also explains this expulsion of the $m_S=0$ states from the central part of the spectrum. 
Also, for $\theta=\pi/2$, there is perfect matching in the behavior of `$-1$' and `$1$' spin projections, which cannot be separated for any energy range. Hence, this situation appears significantly peculiar, since we have no possibility for polarization of `$1$' or `$-1$' currents, although we have transport of `$0$' states in the energy regimes that were previously reserved for `$1$' and `$-1$' states.
We expect similar kind of differences between integer and half-integer spin cases when studying higher spin $S$ states \cite{Pal2016e}.

\subsection*{\label{sec:spinspiral}The spin spiral}

It has been recently shown\cite{Pal2016f} that the choice $\theta_n= 2 \pi n P /L$ --- a spin \emph{spiral} --- for some finite $L$ and $P=2$ leads to a spin flip in the transport for clean systems: at $E<0$ for spin-$1/2$ and $h=3$, initial states with spin-$\uparrow$ emerge after transmitting through the finite system as spin-$\downarrow$ while for $E>0$, spin-$\downarrow$ states emerge as spin-$\uparrow$. Other states are not transmitting. 
On the other hand, for $P=1$, there is no flip and for $E<0$ ($E>0$) only spin-$\uparrow$ (spin-$\downarrow$) states transmit.

\begin{figure}[tb]
\begin{center}
\includegraphics[width=\columnwidth]{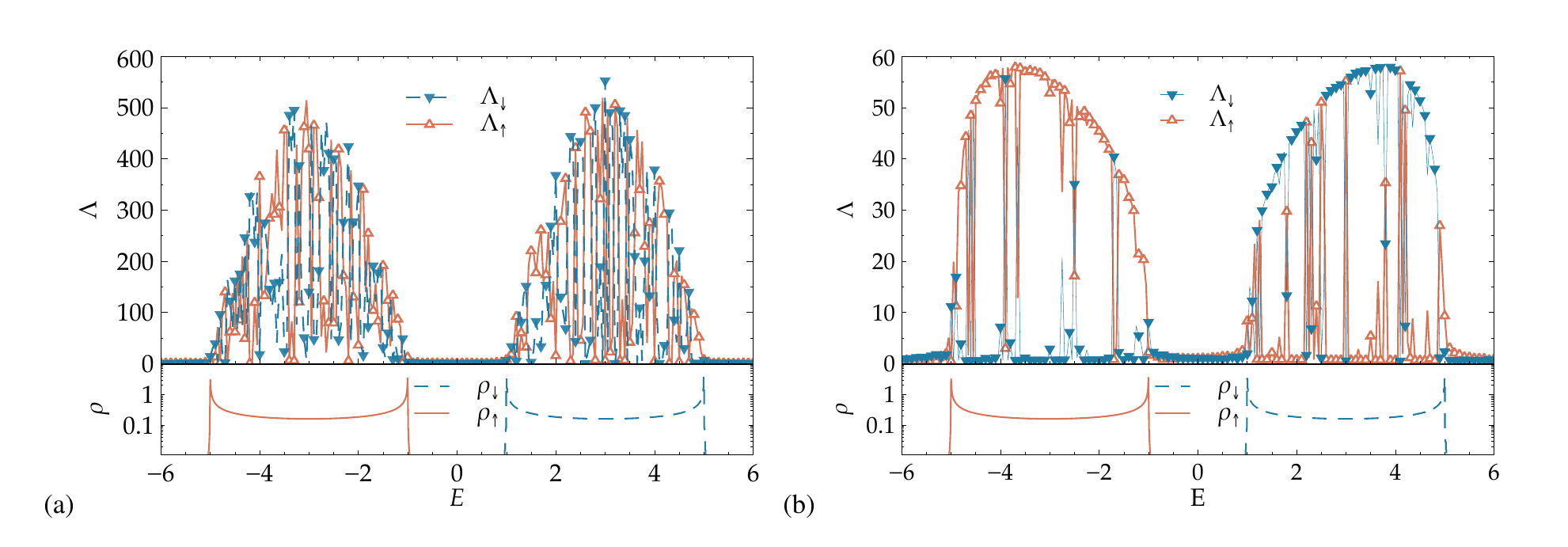} 
\caption{(Color online) Projected localization lengths $\Lambda_{\uparrow}$, $\Lambda_{\downarrow}$ and DOS for the spin spiral case with $W=0.5$ for (a) $P=1$ and (b) $P=2$ at $h=3$. Only every second $\Lambda_\sigma$ symbol is shown for clarity, lines are guides to the eye.}
\label{fig-spinspiral}
\end{center}
\end{figure}

In Fig.\ \ref{fig-spinspiral} we show the analogous situation for finite disorder $W=0.5$. It is immediately clear from the figure that the spin projection fails to separate the spin-$\uparrow$ and spin-$\downarrow$ projections in the case of the spiral (see also Methods). Nevertheless, we still see that the overall localization lengths for $P=1$, cp.\ Fig.\ \ref{fig-spinspiral} (a), are compatible with previous findings\cite{Pal2016f} as before. In particular, the range of energies with finite $\Lambda_\sigma$ values is in excellent agreement. 
For $P=2$, however, the $\Lambda_\sigma$ values are one order of magnitude lower than for $P=1$. This indicates that the flip of spin directions is not captured well by the spin-resolved $\Lambda_\sigma$'s although the reduction in their values might correctly suggest that direct channels ``spin-$\uparrow$ to spin-$\uparrow$'' (similarly for spin-$\downarrow$) are indeed suppressed for $P=2$.
%

\subsection*{\label{sec:spinhalf-magnetic-theta}Disorder in magnetic strength and polar angle }

\begin{figure}[tb]
\begin{center}
\includegraphics[width=\columnwidth]{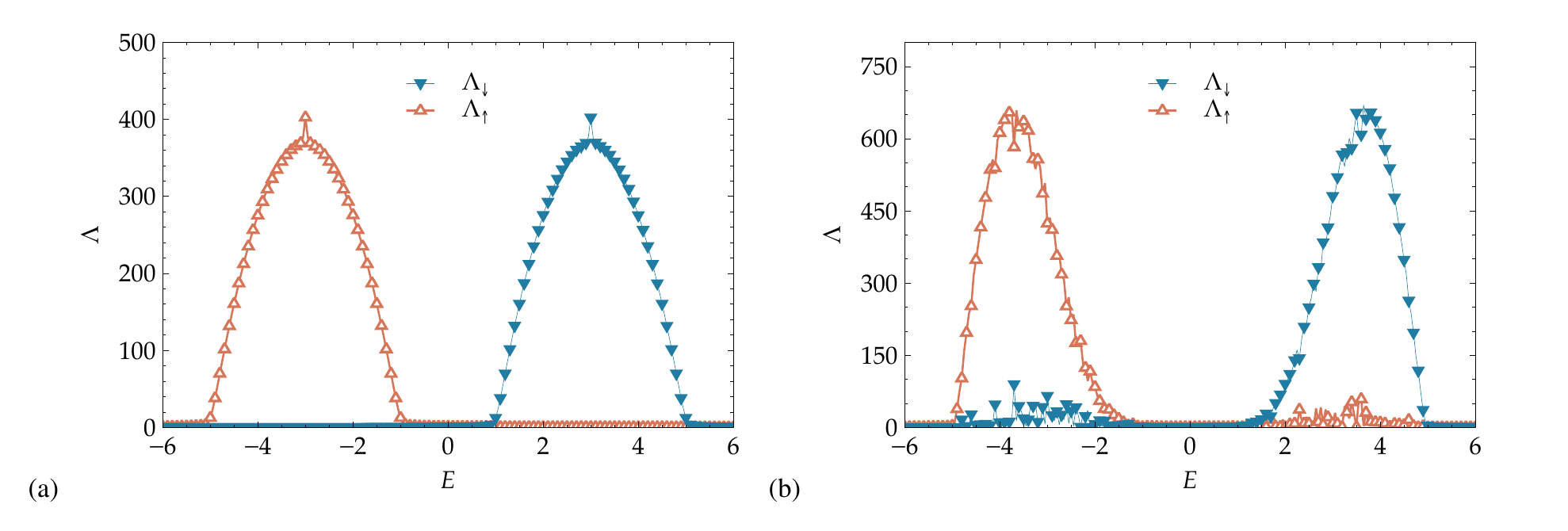}
\caption{(Color online) Projected localization lengths $\Lambda_{\uparrow}$, $\Lambda_{\downarrow}$ with (a) additional magnetic strength disorder $W_h=0.5$ and (b) additional magnetic polar angle disorder $W_{\theta}=0.25$. In both cases, $W=0.1$ and $h=3$. Only every second $\Lambda_\sigma$ symbol is shown for clarity, lines are guides to the eye.} 
\label{fig-spinhalf-magnetic-theta}
\end{center}
\end{figure}

Let us also consider other possible sources of disorder to affect the localization and hence spin filtering properties of the system proposed in Fig.\ \ref{fig-schematic}. In Fig.\ \ref{fig-spinhalf-magnetic-theta} we show results for the spin-$1/2$ case for (a) disorder in $h$ such that at each site $n$, we have $h_n \in [-W_h/2+h, h+W_h/2]$ and (b) disorder in $\theta$ with local value $\theta_n \in [-\theta_h/2+\theta, \theta+\theta_h/2]$. Clearly, these situations correspond to (a) changing magnetic field strength or (b) changing field direction. As is clear from Fig.\ \ref{fig-spinhalf-magnetic-theta}, at the chosen values $W_h=0.5$ or $W_\theta=0.25$, while having a small overall disorder $W=0.1$, the spin-filtering properties remain as robust as in the previously considered cases of pure onsite disorder with $W=0.5$. 

\section*{\label{sec:conclusions}Discussion}
In this article, we have investigated the robustness of spin-polarized transport through a linear array of magnetic atoms depicted in a random landscape.  Flavor of disorder is introduced in the distributions of the on site potential and in the local magnetic field - in its strength as well as in direction. Band gaps are engineered by controlling the strength of the field, and spin-resolved localization lengths are estimated in every case. We developed a new spin projection transfer matrix method for this purpose, which is also applicable to other multi-channel systems. The calculated localization lengths corroborate the observation of spin filtering in every subband, and spin filtering is found to be robust against disorder. 
The results may be inspiring in experimental realizations in the arena of spintronic devices. The unexpected energy selectivity for the dominant spin projection in the spin-$1$ case shows that such studies may indeed be particularly interesting for higher spin cases.
As the basic difference equations hold equally good for bosons, a parallel experimental study in the field of polarized photonic band gap systems may be just on the cards as well \cite{Mukherjee2015}.

\section*{\label{sec:projection}Methods}
\begin{figure}[tb]
\begin{center}
\includegraphics[width=\columnwidth]{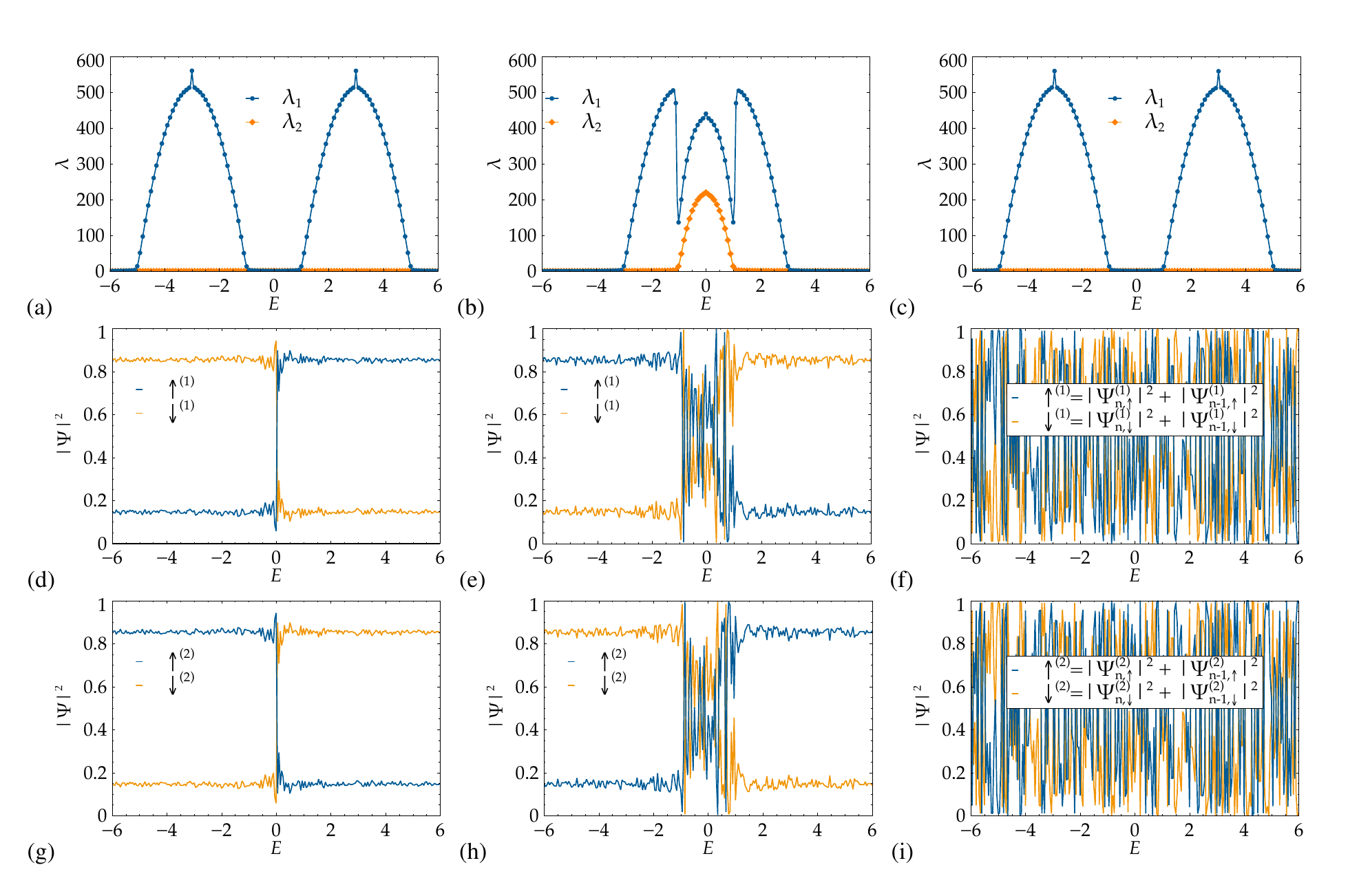} 
\caption{%
(Color online)
First column (a+d+g): (a) Bare localization length $\lambda_1$ and $\lambda_2$ from the TMM for $h=3$ and $\theta=\pi/4$ as in Fig.\ \ref{fig-spinhalf-locpsi-dos} (d). (d+g) Probability amplitudes $\uparrow^{(i)}=\left|\psi^{(i)}_{n,\uparrow}\right|^2+\left|\psi^{(i)}_{n-1,\uparrow}\right|^2$, $\downarrow^{(i)}=\left|\psi^{(i)}_{n,\downarrow}\right|^2+\left|\psi^{(i)}_{n-1,\downarrow}\right|^2$ with $(i=1,2)$ associated with the two transport channels denoted as `1' and `2' for the spin-1/2 case. The results are for $h=3$ and a polar angle $\theta=\pi/4$. 
\label{fig-BARE}
Second column (b+e+h): (b) Bare localization length $\lambda_1$ and $\lambda_2$ for $h=1$ and $\theta=\pi/4$ as in Fig.\ \ref{fig-spinhalf-locpsi-dos} (c). (e+h) Probability amplitudes $\uparrow^{(i)}$, $\downarrow^{(i)}$ with $(i=1,2)$ associated with the channels  `1' and `2'.
Third column (c+f+i): (c) Bare $\lambda_1$ and $\lambda_2$ for the spiral case at $h=3$ and $P=1$. (f+i) Probability amplitudes for the channels `1' and `2'.
}
\label{fig-bare-spinhalf-magnetic-theta}
\end{center}
\end{figure}

The TMM amounts to diagonalizing the transfer matrix $\bb{T}_L=\prod_{n=1}^L T_n$ as $\bb{U}^{-1} \bb{T}_L^{\dagger} \bb{T}_L \bb{U} = \diag(e^{-2L/\lambda_1}$, $e^{-2L/\lambda_2}$, $\ldots$, $e^{-2L/\lambda_{2S+1}}$, $e^{2L/ \lambda_{2S+1}}$, $\ldots$, $e^{2L/\lambda_{1}})$ for a suitably large $L$ to achieve the preset accuracy target \cite{Kramer1993Localization:Experiment}. Each $\lambda_{m_S}$, together with its associated eigenvector $\left( \psi^{(m_S)}_{L,-S}, \ldots, \psi^{(m_S)}_{L,S}, \psi^{(m_S)}_{L-1,-S}, \ldots, \psi^{(m_S)}_{L-1,S} \right)$ correspond to a transport channel, but not automatically to a spin projection $\sigma= -S, \ldots, S$. Instead, the eigenstate corresponding to the largest localization length, say $\lambda_1$, changes its spin content depending on parameters such as $E$. 
In Fig.\ \ref{fig-bare-spinhalf-magnetic-theta} (a), we show an example of this behavior for the same situation as in Fig.\ \ref{fig-spinhalf-locpsi-dos} (d). We observe that while $\lambda_1$ clearly identifies the two regions with finite transport, $\lambda_2$ remains nearly zero throughout the energy band. In contrast, the DOS in Fig.\ \ref{fig-spinhalf-locpsi-dos} (d) shows that the two transport regimes consist of a mixture of both spin components for the spin-$1/2$ situation.
From Figs.\ \ref{fig-bare-spinhalf-magnetic-theta} (d+g), we see that suitably chosen spin projections can distinguish the spin content of each subband: in (d), the probability of the $\sigma=1/2$ projection dominates for $E>0$ while the $\sigma=-1/2$ projection is large for $E<0$; in (g) the situation is reversed, the $\sigma=1/2$ projection dominates for $E<0$ while the $\sigma=-1/2$ projection is large for $E>0$. At $E=0$, in both (d) and (g), we find a rapid change.
In both figures, we note that the sum of the probabilities adds to $1$, due to normalization.

The construction of the \emph{spin-resolved localization lengths} proceeds as in Eq.\ \eqref{eq:spinresolvedloclength}. For $S=1/2$, the explicit form is
\begin{subequations}
\begin{align}
\Lambda_{\uparrow}(n) & = \lambda_1(n) \left(\left|\psi^{(1)}_{n,\uparrow}\right|^2 + \left|\psi^{(1)}_{n-1,\uparrow}\right|^2 \right) + \lambda_2(n) \left(\left|\psi^{(2)}_{n,\uparrow}\right|^2 + \left|\psi^{(2)}_{n-1,\uparrow}\right|^2 \right), \\
\Lambda_{\downarrow}(n) & = \lambda_1(n) \left(\left|\psi^{(1)}_{n,\downarrow}\right|^2 + \left|\psi^{(1)}_{n-1,\downarrow}\right|^2 \right) + \lambda_2(n) \left(\left|\psi^{(2)}_{n,\downarrow}\right|^2 + \left|\psi^{(2)}_{n-1,\downarrow}\right|^2 \right).
\end{align}
\end{subequations}
\revision{We note that $\Lambda_\sigma(n) \leq \lambda_{m_S}(n)$ for all $\sigma$ and $m_{S}$ due to the normalization of the $\psi^{(m_S)}_{n,\sigma}$.} The results \revision{for $\Lambda_\sigma(n)$} as shown in Fig.\ \ref{fig-spinhalf-locpsi-dos} (d) --- and the many other figures showing $\Lambda_\sigma$ throughout the paper --- clearly capture the spin content of the situation very well.
We note that while this interpretation of the eigenvectors in terms of spin projections has not been given before, it is very natural for spatial transport channels in more standard situations.\cite{Gonzalez-Santander2013LocalisationFlakes}
For cases with band overlap, as in Fig.\ \ref{fig-bare-spinhalf-magnetic-theta} (b+e+h), the analogue to Fig.\ \ref{fig-spinhalf-locpsi-dos} (d), it is much harder to interpret the result beyond the realization of spin mixing in $-1 \leq E \leq 1$. Still, even in this situation, the projected $\Lambda_\sigma$ results work well for $|E|>1$.
Last, in the case of the spin-flipping as observed in the spin spiral, the results indicate that the projections fail to capture the progressions of spin-$\uparrow$ changing to spin-$\downarrow$ and, vice versa, $\downarrow$ changing to $\uparrow$.

Let us comment on the determination of errors for $\Lambda_\sigma$. For the $\lambda_\sigma$, we proceed as usual,\cite{Kramer1993Localization:Experiment} using the self-averaging of the Lyapunov exponents and the fact that they are Gaussian distributed after sufficiently many transfer matrix multiplications,\cite{Milde2000a} to compute an error of the mean of at most $0.1\%$. The $\Psi_{n,\sigma}$, however, are not necessarily Gaussian distributed --- indeed, they are mostly not --- nor have they ``converged'' when the Lyapunov exponents already have. This is clearly visible in Fig.\ \ref{fig-bare-spinhalf-magnetic-theta} where the spin projections vary, even for panels (d+g), at the $\sim 5\%$ level while the $\leq 0.1\%$ errors for the $\lambda_\sigma$ are well within symbol size. Furthermore, in the overlap regions, cp.\ Fig.\ \ref{fig-bare-spinhalf-magnetic-theta} (e+h), the fluctuations are much larger and represent a systematic source of variation, different from the purely statistical fluctuations of the $\lambda_\sigma$. We therefore show only the errors computed from the $\lambda_\sigma$ in the figures of main sections.
It may be useful in this context to point out that the computation of the $\Lambda_\sigma$ amounts to computing the spatial content of the transport channels for the $\sigma$'s. This is an even more challenging task than computing the spatial position of a localization center, when given a disorder distribution, which in itself is still an unsolved problem. 


\section*{Acknowledgements}
This work has been supported jointly by the UGC, India and the
British Council through UKIERI, Phase III, references
numbers F.\ 184-14/2017(IC) and UKIERI 2016-17-004 in India and the U.K., respectively. L.B.\ and R.A.R.\ thank the University of Kalyani for its hospitality during their stay in India. A.M.\ is thankful to DST, India for an INSPIRE fellowship [IF160437], and both A.M.\ and A.C.\ gratefully acknowledge the hospitality of the University of
Warwick where this work was completed. A. C. acknowledges financial support through the FRPDF grant from Presidency University, Kolkata. The authors would like to thank Edoardo Carnio, Paul Kelly, Antonio Rodriguez Mesas and Biplab Pal for stimulating discussions and suggestions. We thank the Warwick Scientific Computing Research Technology Platform for computing time and support. 

\section*{Author contributions statement}
A.C.\ and R.A.R.\ conceived the study, A.C.\ and A.M.\ computed the DOS results, L.B.\ and R.A.R.\ computed the spin-resolved localization lengths. All authors contributed to writing and reviewing the manuscript. 

\subsection*{Competing interests} The authors declare that they do not have any financial or non-financial competing interests.
\end{document}